\documentclass{article}
\usepackage{makeidx}
\usepackage{graphicx}
\usepackage{amsmath}

\input{tcilatex}

\begin{document}

\title{Evolutionarily Stable Strategies in Quantum Hawk-Dove Game}
\author{Ahmad Nawaz\thanks{%
email: ahmad@ele.qau.edu.pk} and A.H. Toor\thanks{
\ \  \ \  \ \ \ \ \ ahtoor@qau.edu.pk}$^{\dag }$ \\
National Centre for Physics, Quaid-i-Azam University Campus, \\
Islamabad, Pakistan\\
\dag\ Department of Physics, Quaid-i-Azam University, \\
Islamabad 45320, Pakistan}
\maketitle

\begin{abstract}
We quantized the Hawk-Dove game by using the most general form of a pure
initial state to investigate the existence of pure and mixed Evolutionarily
Stable Strategies (ESS). An example is considered to draw a comparison
between classical and quantum version of the game. Our choice of most
general initial quantum state enables us to make the game symmetric or
asymmetric. We show that for a particular set of game parameters where there
exist only mixed ESS in the classical version of the game, however,
quantization allows even a pure strategy to be an ESS for symmetric game in
addition to ixed ESS. On the other hand only pure strategy ESS can exist for
asymmetric quantum version of the Hawk-Dove game.
\end{abstract}

\section{Introduction}

The concept of evolutionarily stable strategies (ESS) was originally
introduced in evolutionary biology \cite{price}. Later it has been
incorporated as a central concept of stable equilibrium in evolutionarily
game theory. Consider a population in which majority is playing a particular
strategy and a very small fraction of this population, called mutants, start
playing a different strategy. If the mutant strategy remains at disadvantage
against the majority strategy then as a consequence mutants strategy
gradually disappears. In such a situation the majority strategy is called an
ESS. The mathematical theory of ESS was developed for game theoretical
analysis of animal conflicts where they are undergoing natural selection.
Interestingly the dynamics of evolution resulting from Darwinian idea of
survival of fittest implies that ESS must remain stable against small
perturbations caused by mutants. Game theory itself derived much from such
analysis of evolutionary mechanism. Our motivation in present paper is to
give a thought to evolution, as concretely described in notion of ESS, in
circumstances when the game played in population becomes quantum in its
setting.

In an interesting development Meyer \cite{meyer} examined the game theory
from quantum mechanical perspective showing that a player having an access
to quantum strategies can enhance his/her payoff with respect to an opponent
who have access to classical strategies only. Later, Eisert et. al. \cite%
{eisert} analyzed a famous game of Prisoner Dilemma in its quantum
mechanical version and showed that the dilemma no more exists in quantum
version of this game. They also successfully constructed a quantum strategy
which always wins over any classical strategy. Marinatto and Weber \cite%
{marin} proposed another interesting quantization scheme for games where
players can implement their `tactics' on an initial strategy by
probabilistic choice of applying the identity operator $I$ and the flip
operator $C$. They applied their scheme to a famous game of Battle of Sexes
and showed that both the players can get better payoff in a quantum version
of this game.

A consideration of ESS in quantum games is presented by Iqbal and Toor \cite%
{ai,ai1}. They analyzed the quantum games of Prisoner's Dilemma and Battle
of Sexes from the point of view of evolutionary stability. Underlying
assumption in their approach is that a population is playing a quantum game
in two player conflict scenario. They showed that evolutionary stability of
Nash equilibria in symmetric as well as asymmetric games can be controlled
by changing initial quantum state. Recently Prisoner Dilemma has also been
investigated for ESS that is played by using Einstein-Podolsky-Rosen type of
setting \cite{azhar1}.

In this paper we study an interesting game from evolutionary biology called
Hawk-Dove game. In this game Hawk and Dove are the strategies available to
the players to get some valuable resource. The most interesting feature of
this game in its classical form, which is symmetric, is that neither Hawk
nor Dove strategy can be a pure ESS, however, there can exist a mixed ESS.
We have used a most general initial quantum state in quantization that
enables us to make the game symmetric or asymmetric. Our main result in this
paper is that quantization of the game allows even a pure strategy to be an
ESS for symmetric game in addition to mixed ESS. On the other hand only pure
strategy ESS can exist for asymmetric quantum version of the Hawk-Dove game.

The rest of the paper is organized as follows: in Sect. 2 ESS are discussed
from mathematical point of view. Section 3 deals with classical Hawk-Dove
game while in Sect. 4 we present our main result.

\section{Evolutionarily Stable Strategies}

Consider a large population in which members are randomly paired to contest
against each other in a game. In this pair wise contest the average payoff
for a group of members playing strategy $A$ against a small fraction, $%
\epsilon $, of the total population playing strategy $B$ is $%
\$[A,(1-\epsilon )A+\epsilon B].$ Mathematically a strategy $A$ will be an
ESS against a strategy $B,$ if 
\begin{equation}
\$[A,(1-\epsilon )A+\epsilon B]>\$[B,(1-\epsilon )A+\epsilon B],
\label{ESS-defination}
\end{equation}%
there exists a sufficiently small but positive $\epsilon \in \lbrack
0,\epsilon _{0}]$ \cite{broom}. Here $\epsilon _{0}$ is called the invasion
barrier. The strategy played by the smaller group is generally called 
\textit{mutant} strategy and in our case if the fraction of members playing
strategy $B$ becomes larger than $\epsilon _{0},$ then the corresponding
mutant strategy $B$ would be able to invade and strategy $A$ would no longer
be an ESS. In the case of symmetric bi-matrix game between the pairs of
members, the above condition for ESS becomes \cite{broom,prest}:%
\begin{equation*}
\$(A,A)>\$(B,A)
\end{equation*}%
and if $\$(A,A)=\$(B,A)$ then\ $\$(A,B)>\$(B,B)$. For asymmetric case Nash
equilibrium with strict inequality must hold to ensure that strategy $A$ is
an ESS \cite{lann}. For example, a strategy pair $(A^{\ast },B^{\ast })$ is
an ESS if $\$(A^{\ast },B^{\ast })>\$(A,B^{\ast })$ for all $A^{\ast }\neq A$
and $\$_{B}(A^{\ast },B^{\ast })>\$_{B}(A^{\ast },B)\ $for all $B\neq
B^{\ast }.$

\section{Classical Hawk-Dove Game}

Hawk-Dove is a simple two player game where two different behavioral
strategies are available to the players to obtain some resources \cite{keth}%
. Hawks are very aggressive and always fight to take possession of resource.
These fights are very brutal and loser is the one who first sustains the
injury. The winner takes the sole possession of the resource. Mathematical
description of the game requires that the Hawks fully recover before the
next contest. In case both the players opt for Hawk, then the winning
probability for both is equal.

On the other hand Doves never fights for a resource. It displays and if
attacked it immediately withdraws to avoid injury. Thus it will always lose
a conflict against Hawk without sustaining any injury. In other words Doves
fitness remains unaffected. In case two Doves face each other, there will be
period of displaying with some cost (time, energy for display) to both but
without any injury. It is assumed that both the Doves are equally good in
displaying and waiting for random time. In the Dove-Dove contest, both have
equal chance of winning. The winner would be the one with more patience \cite%
{keth}.

Let $\upsilon $ and $i$ are the value of resource and cost of injury,
respectively. Cost of losing a resource is $0,$ while the cost of displaying
is $d$. For mathematical description of the game we assume that $\upsilon $
is a positive number and both $i$ and $d$ are negative numbers. The payoff
matrix for the game takes the form:

\begin{eqnarray}
&&%
\begin{array}{ccc}
\text{ \ \ \ \ \ \ \ \ \ \ \ \ \ \ \ \ \ \ \ \ \ \ \ } & \text{Hawk }(H)%
\text{ \ } & \text{Dove }(D)%
\end{array}
\notag \\
&&%
\begin{array}{c}
\text{Hawk }(H) \\ 
\text{Dove }(D)%
\end{array}%
\left[ 
\begin{array}{cc}
(\frac{\upsilon }{2}+\frac{i}{2},\frac{\upsilon }{2}+\frac{i}{2}) & 
(\upsilon ,0) \\ 
(0,\upsilon ) & (\frac{\upsilon }{2}+d,\frac{\upsilon }{2}+d)%
\end{array}%
\right]  \label{PayMat}
\end{eqnarray}%
It is straight forward to conclude from the above payoff matrix that
strategy Hawk is an ESS if either 
\begin{equation}
\$(H,H)>\$(D,H)  \label{condition-1}
\end{equation}%
or 
\begin{equation}
\$(H,H)=\$(D,H)\ \text{and }\$(H,D)>\$(D,D)  \label{condition-2}
\end{equation}%
For the above payoff matrix (\ref{PayMat}), the first condition translates
to $(\upsilon +i)>0$ and the second to $\upsilon +i=0$ and$\ \frac{\upsilon 
}{2}>d$. Since $\frac{\upsilon }{2}>d$ always holds, therefore, Hawk is an
ESS whenever $\upsilon +i\geq 0.$ Moreover, it important to note that $%
\$(H,D)>\$(D,D)$, hence strategy Dove can never be an ESS \cite{keth}. To
illustrate our point, lets consider the following example where $(\upsilon
+i)<0$ \cite{keth};

\begin{eqnarray}
\text{Value of resource }\upsilon &=&50  \notag \\
\text{Injury to self }i &=&-100  \notag \\
\text{Cost of display }d &=&-10  \notag \\
\text{Resource cost }c &=&0  \label{Example-value}
\end{eqnarray}%
and the payoff matrix (\ref{PayMat}) takes the form

\begin{eqnarray}
&&%
\begin{array}{ccc}
& \text{ \ \ \ \ \ \ \ \ }H\text{ \ \ \ \ \ \ \ } & \text{ \ \ \ \ \ \ \ \ }D%
\end{array}
\notag \\
&&%
\begin{array}{c}
H \\ 
D%
\end{array}%
\left[ 
\begin{array}{cc}
(-25,-25) & (50,0) \\ 
(0,50) & (15,15)%
\end{array}%
\right]  \label{payoffMatrix1}
\end{eqnarray}%
It is clear that there is no ESS for any pure strategy in this game \cite%
{keth}. Next we look at the possibility of ESS in mixed strategies.

Again consider a large population where strategy $H$ is being played by a
fraction $h$ of the total population and remaining population is playing
strategy $D.$ In a pair wise contest the corresponding fitnesses functions $%
W(H)$, $W(D)$ are defined as \cite{keth}

\begin{subequations}
\begin{eqnarray}
W(H) &=&\$(H,H)h+\$(H,D)(1-h),  \notag \\
W(D) &=&\$(D,H)h+\$(D,D)(1-h).  \label{fit}
\end{eqnarray}%
For a mixed strategy to be an ESS, for both the strategies we must have
equal fitness functions, which implies

\end{subequations}
\begin{equation}
h=\frac{2d-\upsilon }{2d+i}.  \label{fitness}
\end{equation}%
Lets explore the possibility of mixed strategy ESS for the set of values
considered in the above example [see Eq. (\ref{Example-value})]. Putting
these values in the Eq. (\ref{fitness}), we see that there exist an ESS for $%
h=.583.$

\section{Quantum Hawk-Dove Game}

We follow the Marinatto and Weber's scheme to quantize the strategy space
for Hawk and Dove game \cite{marin}. Assuming that two players, Alice and
Bob, share the following entangled state:

\begin{equation}
\left\vert \psi _{in}\right\rangle =a\left\vert HH\right\rangle +b\left\vert
DD\right\rangle +c\left\vert HD\right\rangle +d\left\vert DH\right\rangle
\label{enst2}
\end{equation}%
where $\left\vert a\right\vert ^{2}+\left\vert b\right\vert ^{2}+\left\vert
c\right\vert ^{2}+\left\vert d\right\vert ^{2}=1$ and the first slot is
reserved for Alice's strategy and the second for Bob's. Let $C$ be a unitary
and Hermitian operator (i.e., $C=C^{\dagger }=C^{-1}$), such that 
\begin{equation}
C\left\vert H\right\rangle =\left\vert D\right\rangle \text{, }C\left\vert
D\right\rangle =\left\vert H\right\rangle \text{. }  \label{oper}
\end{equation}%
If Alice uses $I$, the identity operator, with probability $p$ and $C$ with
probability $(1-p)$ and Bob uses these operators with probability $q$ and $%
(1-q)$, respectively. Then the final density matrix of the bipartite system
takes the form \cite{mtb};%
\begin{eqnarray}
\rho _{f} &=&pq\left[ \left( I_{A}\otimes I_{B}\right) \rho _{in}\left(
I_{A}^{\dagger }\otimes I_{B}^{\dagger }\right) \right]  \notag \\
&&+p(1-q)\left[ \left( I_{A}\otimes C_{B}\right) \rho _{in}\left(
I_{A}^{\dagger }\otimes C_{B}^{\dagger }\right) \right]  \notag \\
&&+q(1-p)\left[ \left( C_{A}\otimes I_{B}\right) \rho _{in}\left(
C_{A}^{\dagger }\otimes I_{B}^{\dagger }\right) \right]  \notag \\
&&+(1-p)(1-q)\left[ \left( C_{A}\otimes C_{B}\right) \rho _{in}\left(
C_{A}^{\dagger }\otimes C_{B}^{\dagger }\right) \right] .  \label{def}
\end{eqnarray}%
Here $\rho _{in}=\left\vert \psi _{in}\right\rangle \left\langle \psi
_{in}\right\vert .$ The payoff operators for Alice and Bob are defined as %
\cite{mtb}

\begin{eqnarray}
P_{A} &=&(\frac{\upsilon }{2}+\frac{i}{2})\left\vert HH\right\rangle
\left\langle HH\right\vert +\upsilon \left\vert HD\right\rangle \left\langle
HD\right\vert +(\frac{\upsilon }{2}+d)\left\vert DD\right\rangle
\left\langle DD\right\vert ,  \label{popa} \\
P_{B} &=&(\frac{\upsilon }{2}+\frac{i}{2})\left\vert HH\right\rangle
\left\langle HH\right\vert +\upsilon \left\vert DH\right\rangle \left\langle
DH\right\vert +(\frac{\upsilon }{2}+d)\left\vert DD\right\rangle
\left\langle DD\right\vert .  \label{popb}
\end{eqnarray}%
The payoff functions for Alice and Bob are the mean values of the above
operators, i.e.,

\begin{eqnarray}
\$_{A}(p,q) &=&\text{Tr}(P_{A}\rho _{f}),\text{ }  \notag \\
\text{\ \ \ \ }\$_{B}(p,q) &=&\text{Tr}(P_{B}\rho _{f}).  \label{payf}
\end{eqnarray}%
The expected payoff functions for both the players are obtained using Eqs. (%
\ref{def},\ref{payf})%
\begin{eqnarray}
\$_{A}(p,q) &=&(\frac{\upsilon }{2}+\frac{i}{2})[pq\left\vert a\right\vert
^{2}+p(1-q)\left\vert c\right\vert ^{2}+q(1-p)\left\vert d\right\vert
^{2}+(1-p)(1-q)\left\vert b\right\vert ^{2}]  \notag \\
&&+\upsilon \lbrack pq\left\vert c\right\vert ^{2}+p(1-q)\left\vert
a\right\vert ^{2}+q(1-p)\left\vert b\right\vert ^{2}+(1-p)(1-q)\left\vert
d\right\vert ^{2}]  \notag \\
&&+(\frac{\upsilon }{2}+d)[pq\left\vert b\right\vert ^{2}+p(1-q)\left\vert
d\right\vert ^{2}+q(1-p)\left\vert c\right\vert ^{2}+(1-p)(1-q)\left\vert
a\right\vert ^{2},  \notag \\
\$_{B}(p,q) &=&(\frac{\upsilon }{2}+\frac{i}{2})[pq\left\vert a\right\vert
^{2}+p(1-q)\left\vert c\right\vert ^{2}+q(1-p)\left\vert d\right\vert
^{2}+(1-p)(1-q)\left\vert b\right\vert ^{2}]  \notag \\
&&+\upsilon \left[ pq\left\vert d\right\vert ^{2}+p(1-q)\left\vert
b\right\vert ^{2}+q(1-p)\left\vert a\right\vert ^{2}+(1-p)(1-q)\left\vert
c\right\vert ^{2}\right]  \notag \\
&&+(\frac{\upsilon }{2}+d)[pq\left\vert b\right\vert ^{2}+p(1-q)\left\vert
d\right\vert ^{2}+q(1-p)\left\vert c\right\vert ^{2}+(1-p)(1-q)\left\vert
a\right\vert ^{2}].  \notag \\
&&  \label{pay1}
\end{eqnarray}%
Corresponding to the set of values we considered earlier, i.e., Eq. (\ref%
{payoffMatrix1}), the above set of payoff functions becomes: 
\begin{eqnarray}
\$_{A}(p,q) &=&p[q\{-60\left\vert a\right\vert ^{2}-60\left\vert
b\right\vert ^{2}+60\left\vert c\right\vert ^{2}+60\left\vert d\right\vert
^{2}\}  \notag \\
&&-25\left\vert c\right\vert ^{2}+25\left\vert b\right\vert
^{2}+35\left\vert a\right\vert ^{2}-35\left\vert d\right\vert ^{2}]  \notag
\\
&&+q[75\left\vert b\right\vert ^{2}-75\left\vert d\right\vert
^{2}-15\left\vert a\right\vert ^{2}+15\left\vert c\right\vert ^{2}]  \notag
\\
&&-25\left\vert b\right\vert ^{2}+50\left\vert d\right\vert
^{2}+15\left\vert a\right\vert ^{2},  \notag \\
\$_{B}(p,q) &=&q[p\{-60\left\vert a\right\vert ^{2}-60\left\vert
b\right\vert ^{2}+60\left\vert c\right\vert ^{2}+60\left\vert d\right\vert
^{2}\}  \notag \\
&&-25\left\vert d\right\vert ^{2}+25\left\vert b\right\vert
^{2}+35\left\vert a\right\vert ^{2}-35\left\vert c\right\vert ^{2}]  \notag
\\
&&+p[75\left\vert b\right\vert ^{2}-75\left\vert c\right\vert
^{2}+15\left\vert d\right\vert ^{2}-15\left\vert a\right\vert ^{2}]  \notag
\\
&&-25\left\vert b\right\vert ^{2}+50\left\vert c\right\vert
^{2}+15\left\vert a\right\vert ^{2}.  \label{pay2}
\end{eqnarray}%
As the classical version of the Hawk and Dove game is symmetric, one would
expect in the quantum version of the game interchanging $p$ and $q$ would
change $\$_{A}(p,q)$ into $\$_{B}(p,q)$. However, it is interesting to note
that in quantum version of the game would be symmetric, if $c=d$ in initial
quantum state $\left\vert \psi _{in}\right\rangle $ (\ref{enst2}). This
observation is consistent with earlier work on {\normalsize ESS} where
quantum version of the game is shown to be symmetric \cite{azhar,ai,ai1}. In
our case, the possibility of asymmetric game is due to the choice of general
initial quantum state instead of one of the Bell states. Next we discuss
both symmetric and asymmetric case in our game separately.\textrm{\ }

\subsection{Symmetric case}

Our generalized treatment of Hawk and Dove game become symmetric for game
for $c=d$ in initial state. The corresponding payoff functions, i.e., Eq. (%
\ref{pay2}), become:

\begin{eqnarray}
\$_{A}(p,q) &=&p[q\{-60\left\vert a\right\vert ^{2}-60\left\vert
b\right\vert ^{2}+120\left\vert c\right\vert ^{2}\}+35\left\vert
a\right\vert ^{2}+25\left\vert b\right\vert ^{2}-60\left\vert c\right\vert
^{2}]  \notag \\
&&+q[-15\left\vert a\right\vert ^{2}+75\left\vert b\right\vert
^{2}-60\left\vert c\right\vert ^{2}]+15\left\vert a\right\vert
^{2}-25\left\vert b\right\vert ^{2}+50\left\vert c\right\vert ^{2},  \notag
\\
\$_{B}(p,q) &=&q[p\{-60\left\vert a\right\vert ^{2}-60\left\vert
b\right\vert ^{2}+120\left\vert c\right\vert ^{2}\}+35\left\vert
a\right\vert ^{2}+25\left\vert b\right\vert ^{2}-60\left\vert c\right\vert
^{2}]  \notag \\
&&+p[-15\left\vert a\right\vert ^{2}+75\left\vert b\right\vert
^{2}-60\left\vert c\right\vert ^{2}]+15\left\vert a\right\vert
^{2}-25\left\vert b\right\vert ^{2}+50\left\vert c\right\vert ^{2}.
\label{sym}
\end{eqnarray}%
Being a symmetric game the players are anonymous, therefore the subscripts $%
A $ and $B$ are not necessary, i.e., $\$_{A}(p,q)=\$_{B}(p,q)=\$(p,q)$.
Corresponding NE inequality becomes%
\begin{equation*}
\$(p^{\ast },q^{\ast })-\$(p,q^{\ast })\geq 0
\end{equation*}%
This condition in our example translates to:

\begin{equation}
(p^{\ast }-p)[q^{\ast }\{-60\left\vert a\right\vert ^{2}-60\left\vert
b\right\vert ^{2}+120\left\vert c\right\vert ^{2}\}+35\left\vert
a\right\vert ^{2}+25\left\vert b\right\vert ^{2}-60\left\vert c\right\vert
^{2}]\geq 0  \label{nash1}
\end{equation}%
Upon inspection it can be seen that the above inequality holds if both the
factors have same sign. Lets consider following three cases:

\subsubsection{Case 1:}

Lets consider a case of a pure strategy $\left( p^{\ast }=0,q^{\ast
}=0\right) $ and examine the possibility of it being a NE. The inequality (%
\ref{nash1}) for this strategy requires $35\left\vert a\right\vert
^{2}+25\left\vert b\right\vert ^{2}-60\left\vert c\right\vert ^{2}<0$. This
holds, for example, when $\left\vert a\right\vert ^{2}=\frac{1}{16}%
,\left\vert b\right\vert ^{2}=\frac{1}{4},\left\vert c\right\vert ^{2}=\frac{%
11}{32}$. The corresponding payoff functions from Eqs. (\ref{sym}) are

\begin{eqnarray}
\$(0,0) &=&\frac{95}{8},  \notag \\
\$(p,0) &=&\frac{95}{8}-\frac{195}{16}p.
\end{eqnarray}%
Which means $\$(0,0)>\$(p,0)$ $\forall $ $0<p<1$, therefore, the strategy $%
\left( p^{\ast }=0,q^{\ast }=0\right) $ is an ESS. In the context of our
evolutionary game if both the players are playing strategy $C$ no mutant
strategy can invade for an initial state, $\left\vert \psi
_{in}\right\rangle =\frac{1}{4}\left\vert HH\right\rangle +\frac{1}{2}%
\left\vert DD\right\rangle +\sqrt{\frac{11}{32}}\left\vert HD\right\rangle +%
\sqrt{\frac{11}{32}}\left\vert DH\right\rangle .$ Therefore, in contrast to
the classical version of the game pure strategies can also be a ESS in the
quantum version of the game under certain conditions.

\subsubsection{Case 2:}

Lets examine another case of pure strategy $\left( p^{\ast }=1,q^{\ast
}=1\right) $ to be a NE. For this strategy the inequality given by Eq. (\ref%
{nash1}) demands $-25\left| a\right| ^{2}-35\left| b\right| ^{2}+60\left|
c\right| ^{2}>0$. This holds, for example, for $\left| a\right| ^{2}=\frac{1%
}{16},\left| b\right| ^{2}=\frac{1}{8},\left| c\right| ^{2}=\frac{13}{32}$.
The corresponding payoff functions from Eqs. (\ref{sym}) are 
\begin{eqnarray}
\$(1,1) &=&\frac{165}{8}  \notag \\
\$(p,1) &=&\frac{35}{16}+\frac{295}{16}p.
\end{eqnarray}%
Since $\$(1,1)-\$(p,1)=\frac{295}{16}(1-p)>0$ $\forall $ $0<p<1$, therefore, 
$\left( p^{\ast }=1,q^{\ast }=1\right) $ is an ESS. Thus a population
engaged in the pure strategy$\left( p^{\ast }=1,q^{\ast }=1\right) $ cannot
be invaded by any mutant strategy if for the initial quantum state the
inequality $-25\left| a\right| ^{2}-35\left| b\right| ^{2}+60\left| c\right|
^{2}>0$ holds, which correspond to the initial state $\left| \psi
_{in}\right\rangle =\frac{1}{4}\left| HH\right\rangle +\sqrt{\frac{1}{8}}%
\left| DD\right\rangle +\sqrt{\frac{13}{32}}\left| HD\right\rangle +\sqrt{%
\frac{13}{32}}\left| DH\right\rangle $.

\subsubsection{\textbf{Case 3:}}

Lets explore the possibility of mixed NE in the quantum version of the game
when the players apply their operators with probability\textrm{\ }$0<p<1$.
From the inequality given by Eq. (\ref{nash1}),$\ $the mixed NE is 
\begin{equation}
p^{\ast }=q^{\ast }=\frac{-7\left| a\right| ^{2}-5\left| b\right|
^{2}+12\left| c\right| ^{2}}{12(-\left| a\right| ^{2}-\left| b\right|
^{2}+2\left| c\right| ^{2})}.  \label{mixed-p}
\end{equation}%
Corresponding to the classical version of the game, we get $\left( p^{\ast }=%
\frac{7}{12},q^{\ast }=\frac{7}{12}\right) $. In quantum version of the game
we can obtain this value, for example, for $\left| a\right| ^{2}=\frac{1}{2}%
,\left| b\right| ^{2}=\left| c\right| ^{2}=\frac{1}{6}$. The initial state,
then, takes the form 
\begin{equation}
\left| \psi _{in}\right\rangle =\frac{1}{\sqrt{2}}\left| HH\right\rangle +%
\frac{1}{\sqrt{6}}\left| DD\right\rangle +\frac{1}{\sqrt{6}}\left|
HD\right\rangle +\frac{1}{\sqrt{6}}\left| DH\right\rangle  \label{inst}
\end{equation}%
Now from Eq. (\ref{sym}) 
\begin{eqnarray*}
\$(p^{\ast },q^{\ast }) &=&\$(p,q^{\ast })=8.75, \\
\$(q,q) &=&\frac{-60q^{2}+20q+35}{3}, \\
\$(p^{\ast },q) &=&\frac{-600q+665}{36}.
\end{eqnarray*}%
It can be seen that $\$(p^{\ast },q)-\$(q,q)>0$, $\forall $ $0<q<1$. This
implies that $\$(p^{\ast },q)>\$(q,q)$. Therefore $\left( p^{\ast },q^{\ast
}\right) $ given by Eq. (\ref{mixed-p}) is a mixed ESS for the above initial
quantum state.

\subsection{Asymmetric case}

Our initial quantum state correspond to asymmetric game for $c\neq d$ for
which an ESS is defined with strict Nash inequality \cite{lann}. In this
case a strategy pair $(A^{\ast },B^{\ast })$ is an ESS if NE conditions with
strict inequalities hold, i.e.,\ $\$_{A}(A^{\ast },B^{\ast
})>\$_{A}(A,B^{\ast })$\ for all $A\neq A^{\ast }$ and $\$_{B}(A^{\ast
},B^{\ast })>\$_{B}(A^{\ast },B)$ for all $B\neq B^{\ast }$. Nash
inequalities (\ref{pay2}) then yield

\ 
\begin{gather}
\$_{A}(p^{\ast },q^{\ast })-\$_{A}(p,q^{\ast })\geq 0  \notag \\
\Rightarrow (p^{\ast }-p)[60q^{\ast }\{-\left| a\right| ^{2}-\left| b\right|
^{2}+\left| c\right| ^{2}+\left| d\right| ^{2}\}+35\left| a\right|
^{2}+25\left| b\right| ^{2}-25\left| c\right| ^{2}-35\left| d\right|
^{2}]\geq 0  \label{ne1}
\end{gather}%
and%
\begin{gather}
\$_{B}(p^{\ast },q^{\ast })-\$_{A}(p^{\ast },q)\geq 0  \notag \\
\Rightarrow (q^{\ast }-q)[60p^{\ast }\{-\left| a\right| ^{2}-\left| b\right|
^{2}+\left| c\right| ^{2}+\left| d\right| ^{2}\}+35\left| a\right|
^{2}+25\left| b\right| ^{2}-35\left| c\right| ^{2}-25\left| d\right|
^{2}]\geq 0  \label{ne2}
\end{gather}%
From these inequalities three Nash equilibria arise

\subsubsection{Case 1:}

From inequalities (\ref{ne1}), (\ref{ne2}) with $\left( p^{\ast }=0,q^{\ast
}=0\right) ,$ we get

\begin{eqnarray}
35\left\vert a\right\vert ^{2}+25\left\vert b\right\vert ^{2}-25\left\vert
c\right\vert ^{2}-35\left\vert d\right\vert ^{2} &<&0  \label{equ1} \\
35\left\vert a\right\vert ^{2}+25\left\vert b\right\vert ^{2}-35\left\vert
c\right\vert ^{2}-25\left\vert d\right\vert ^{2} &<&0  \label{eq2}
\end{eqnarray}%
respectively. Both these inequalities (\ref{equ1}) and (\ref{eq2}) are
satisfied, for example, for $\left\vert a\right\vert ^{2}=\frac{1}{16}%
,\left\vert b\right\vert ^{2}=\frac{1}{4},\left\vert c\right\vert ^{2}=\frac{%
9}{16},\left\vert d\right\vert ^{2}=\frac{1}{8}$. Therefore, from Eq. (\ref%
{pay2})%
\begin{eqnarray}
\$_{A}(0,0) &=&\frac{15}{16}  \notag \\
\$_{A}(p,0) &=&\frac{15}{16}-10p
\end{eqnarray}%
\begin{eqnarray}
\$_{B}(0,0) &=&\frac{365}{16}  \notag \\
\$_{B}(0,q) &=&\frac{365}{16}-\frac{230}{16}q
\end{eqnarray}%
Since $\$_{A}(0,0)>\$_{A}(p,0)$ $\forall $ $0<p<1$ and $\$_{B}(0,0)>%
\$_{B}(0,q)$ $\forall $ $0<q<1.$ Therefore, strict inequality holds and
strategy $\left( p^{\ast }=0,q^{\ast }=0\right) $ is an ESS.

\subsubsection{Case 2:}

Similarly from inequalities (\ref{ne1}), (\ref{ne2}) with\ $\left( p^{\ast
}=1,q^{\ast }=1\right) ,$ we get

\begin{eqnarray}
-25\left| a\right| ^{2}-35\left| b\right| ^{2}+35\left| c\right|
^{2}+25\left| d\right| ^{2} &>&0  \label{eq3} \\
-25\left| a\right| ^{2}-35\left| b\right| ^{2}+25\left| c\right|
^{2}+35\left| d\right| ^{2} &>&0  \label{eq4}
\end{eqnarray}%
respectively. These inequalities are satisfied, for example, for $\left|
a\right| ^{2}=\frac{1}{16}$, $\left| b\right| ^{2}=\frac{1}{8}$, $\left|
c\right| ^{2}=\frac{9}{16}$, $\left| d\right| ^{2}=\frac{1}{4}$. Hence from
eq. (\ref{pay2}) 
\begin{align}
\$_{A}(1,1)& =\frac{455}{16}  \notag \\
\$_{A}(p,1)& =\frac{135}{16}+20p
\end{align}%
\begin{align}
\$_{B}(1,1)& =\frac{205}{16}  \notag \\
\$_{B}(1,q)& =\frac{270}{16}q-\frac{65}{16}
\end{align}%
Again it shows that $\$_{A}(1,1)>\$_{A}(p,1)$ $\forall $ $0<p<1$ and $%
\$_{B}(1,1)>\$_{B}(1,q)$ $\forall $ $0<q<1$. As strict inequality holds in
this case, therefore,$\left( p^{\ast }=q^{\ast }=1\right) $ is an ESS.

\subsubsection{\textbf{Case 3:}}

For asymmetric case from inequalities (\ref{ne1}), (\ref{ne2}), we get

\begin{eqnarray}
p^{\ast } &=&\frac{-7\left\vert a\right\vert ^{2}-5\left\vert b\right\vert
^{2}+7\left\vert c\right\vert ^{2}+5\left\vert d\right\vert ^{2}}{%
12(-\left\vert a\right\vert ^{2}-\left\vert b\right\vert ^{2}+\left\vert
c\right\vert ^{2}+\left\vert d\right\vert ^{2})},  \label{mixed} \\
q^{\ast } &=&\frac{-7\left\vert a\right\vert ^{2}-5\left\vert b\right\vert
^{2}+5\left\vert c\right\vert ^{2}+7\left\vert d\right\vert ^{2}}{%
12(-\left\vert a\right\vert ^{2}-\left\vert b\right\vert ^{2}+\left\vert
c\right\vert ^{2}+\left\vert d\right\vert ^{2})}.
\end{eqnarray}%
Strict inequality does not hold for these values, therefore, it is not an
ESS.

\section{Summary}

Evolutionary game theory with ESS as central idea is an interesting branch
of game theory. It was developed by mathematical biologists to model
evolutionary dynamics. Introduction of quantum mechanics in evolutionary
game theory transpired very interesting situations, e.g., in a quantum
version of Rock Scissor Paper (RSP) mixed NE becomes stable contrary to
classical version of the game where no stable mixed NE exit \cite{azhar}.
Similarly quantization of the Prisoner Dilemma and the Battle of Sexes
showed that evolutionary stability of NE in symmetric as well in asymmetric
games can be changed by maneuvering initial quantum state \cite{ai,ai1}.

We quantized the Hawk-Dove game using a pure initial quantum state of
two-qubit system in its most general form. We showed that for quantization
of this symmetric classical game, initial quantum state plays a crucial role
in keeping it symmetric or asymmetric. In other words there is a restriction
on initial quantum state for which a classical game remains symmetric in its
quantum form. To elaborate our point we considered an example with set of
parameters for which there is no pure ESS for the classical Hawk-Dove game
though there exits mixed ESS. However in quantum version of the game, even a
pure strategy can be an ESS for certain initial quantum state. We analyzed
both symmetric and asymmetric situations and showed that pure ESS can exist
in both symmetric and asymmetric whereas mixed ESS exists only in the
symmetric form of the quantum version of the game.

\section{Acknowledgment}

One of us (A. N) is grateful to A. Iqbal and Khalid Loan for their useful
help.

\end{document}